\documentclass[twocolumn,floatfix,amsmath,amssymb,aps]{revtex4}
\usepackage{epsfig}
\begin{document}
\title{Critical dynamics of diluted relaxational models coupled to a conserved density (diluted model C)}
\author{M.~Dudka}
\affiliation{ Institute for Condensed Matter Physics,
National Academy of Sciences of Ukraine, UA-79011 Lviv, Ukraine}
\affiliation{Institute for Theoretical Physics, University of Linz, A-4040 Linz, Austria}
\author{R.~Folk}
\affiliation{Institute for Theoretical Physics, University of Linz, A-4040 Linz, Austria}
\author{Yu.~Holovatch}
\affiliation{ Institute for Condensed Matter Physics,
National Academy of Sciences of Ukraine, UA-79011 Lviv, Ukraine}
\affiliation{Institute for Theoretical Physics, University of Linz,A-4040 Linz, Austria}
\affiliation{Ivan Franko National University of Lviv, UA-79005 Lviv, Ukraine}
\author{G.~Moser}
\affiliation{Institute for Physics and Biophysics, University of Salzburg, A-5020 Salzburg, Austria}

\begin{abstract}
We consider the influence of quenched disorder on the relaxational critical dynamics of a
system characterized by a non-conserved order parameter coupled to the diffusive
dynamics of a conserved scalar density (model C). Disorder leads to model A
critical dynamics in the asymptotics, however it is the effective critical behavior
which is often observed in experiments and in computer simulations and this is described by
the full set of dynamical equations of diluted model C. Indeed different scenarios of effective critical behavior
are predicted.

\vspace{6pt}
\noindent Pacs numbers: 05.70.Jk; 64.60.Ht; 64.60.Ak

\end{abstract}

\maketitle

The critical behavior of pure systems might be changed by introducing imperfections
like dilution, defects, etc. into a critical system. If such a change can  be
expected is answered by the Harris criterion \cite{Harris74} stating that a new diluted critical
behavior appears if the specific heat of the pure system is diverging. The diluted
critical behavior then has a nondiverging specific heat. Since the borderline value $n_c$
between a diverging and nondiverging specific heat at space dimensions $d=3$ lies
between order parameter (OP) dimensions $n=1$ (Ising model)
and $n=2$ (XY model) only the Ising case belongs to a new universality class.
In consequence this result
led to the conclusion that for the critical {\bf dynamics} the coupling of
conserved quantities to the OP is of no relevance \cite{Krey77,Lawrie84}.
The argument was the following: For the critical dynamics of a relaxational model it
was shown \cite{hahoma74,FoMo03,Folk04} that the coupling to
a conserved density (e.g. the energy density) is relevant if the specific heat
diverges. Due to dilution this is never the case and therefore the coupling is of no relevance.
Therefore most of the papers considered only the relaxational dynamics of Ising systems
\cite{Grinstein77,Prudnikov92,Oerding95,Janssen95}

However this argumentation is based on the {\bf asymptotic} properties of the
diluted model. Experimental data and computer simulations made clear that
in most cases one observes {\bf non-asymptotic} critical behavior, described often by
dilution dependent effective exponents (see e.g. \cite{review,perumal03}). In such a case the Harris criterion
does not hold and therefore one has to consider in the dynamics the coupling to the
conserved density and its effects on the effective critical behavior. In addition
one is not restricted to the Ising case since already in statics the effective
critical behavior for $n>1$ is different from the pure case \cite{perumal03}.

There are two relevant  parameters of model C: (i) the static coupling $\gamma$ of the OP to the conserved
density and (ii) a dynamic parameter, the time scale ratio $w=\Gamma/\lambda$  where $\Gamma$ is the relaxation
rate of the OP and $\lambda$ is the diffusion rate of the conserved density.
From the renormalization group (RG) treatment of model C one knows that the one loop order does not give reliable results due to
the stability of a fixed point with the time scale ratio  $w=\infty$. In two loop (and higher) order it turns out
that this fixed point is unstable and model C is characterized by strong and weak scaling regions for the dynamics at $d=3$
\cite{FoMo03,Folk04}. Moreover it was shown that non-asymptotic effects are already present in model C \cite{Folk04}.
In the following we will consider how these aspects are influenced by disorder.

Model C \cite{hahoma74,FoMo03} describes the relaxational dynamics of a system characterized by
an $n$-component nonconserved OP $\vec{\varphi}_0(x,t)$ coupled to the diffusive dynamics of
a  conserved  scalar density $m_0(x,t)$. The structure of
the equations of motions is not changed by the presence of disorder. They read:
\begin{equation}\label{eq_mov2}
\frac{\partial \vec{\varphi}_{0}}{\partial
t}=-\mathring{\Gamma}\frac{\partial {\mathcal H}}{\partial
\vec{\varphi}_{0}}+\vec{\theta}_{{\varphi}} \, , \quad
%\label{eq_mov2v}
 \frac{\partial {m}_0}{\partial
t}=\mathring{\lambda}\nabla^2\frac{\partial {\mathcal
H}}{\partial {m}_0}+{\theta_{{m}}}
\end{equation}
where  $_0$ or $\mathring{}$ denote unrenormalized quantities. The stochastic forces in (\ref{eq_mov2}) satisfy the Einstein relations:
\begin{eqnarray}\label{1}
<{\theta}_{\varphi_i}(x,t){\theta}_{\varphi_j}(x',t')>&=&2\mathring{\Gamma}\delta(x-x')\delta(t-t')\delta_{ij},
\\ \label{2}
 <{\theta}_{m}(x,t){\theta}_{{m}}(x',t')>&=&-2\mathring{\lambda}\nabla^2\delta(x-x')\delta(t-t') \, .
\end{eqnarray}

Equilibrium is described by the static functional $\mathcal H$ of the {\bf disordered} magnetic system
\begin{eqnarray}\label{hamilt1}
{\mathcal H}&=&\int d^d x \Big \{
 \frac{1}{2}\mathring{\tilde{r}} |\vec{\varphi}_0|^2 + V(x)|\vec{\varphi}_0|^2
+\frac{1}{2}\sum_{i=1}^{n}(\nabla \varphi_{i,0})^2     \\
&+&\frac{\mathring{\tilde{u}}}{4!}|\vec{\varphi}_0|^4 +
\frac{1}{2}a_{{m}}{{m_0}}^2+ \frac{1}{2}\mathring{\gamma}
{{m_0}}|\vec{\varphi}_0|^2-\mathring{h}_{{m}}{{m_0}}
 \Big \},  \nonumber
\end{eqnarray}
where $V(x)$ is an impurity potential which introduces disorder to the system, $d$ is the spatial
dimension.
It contains a coupling $\mathring{\gamma}$ to the secondary density which can be
integrated out. Thus  static critical properties described by the functional
(\ref{hamilt1}) are equivalent to those of the functional
${\mathcal H}=\int d^d x \Big \{ \frac{1}{2}\mathring{r} |\vec{\varphi}_0|^2 + V(x)|\vec{\varphi}_0|^2
+\frac{1}{2}\sum_{i=1}^{n}(\nabla \varphi_{i,0})^2 +\frac{\mathring{u}}{4!}|\vec{\varphi}_0|^4\Big \}$.
 The parameters $\mathring{r}$ and $\mathring{u}$ are related to
$ \mathring{\tilde r}$, $\mathring{\tilde u}$, $a_m$, $\mathring{\gamma}$ and $\mathring{h}_m$ by
$\mathring{r}=\mathring{\tilde{r}}+\mathring{\gamma}\mathring{h}_{{m}}/{a_m}$ and
$\mathring{u}=\mathring{\tilde{u}}-3\mathring{\gamma}^2/a_m$.
\begin{figure}  \centering
\epsfig{file=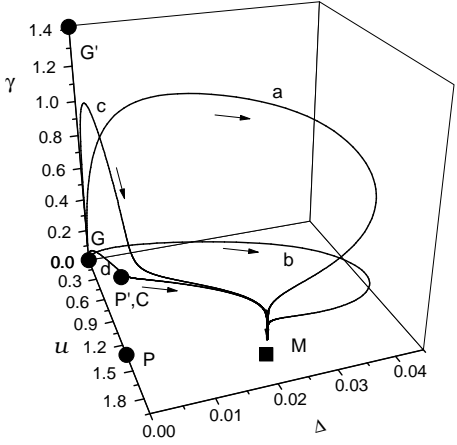,width=6.0cm,angle=0}
\caption{\label{n1statflow} Two loop flow of static parameters for $n=1$.
The stable fixed point (full square) has a finite width $\Delta$ of the impurity distribution
and leads to asymptotic exponents different from the pure case (unstable fixed points are
indicated by full circles). Depending on the initial values of
the renormalized couplings the effective exponents are quite different.}
\end{figure}
$\mathring{r}$ is proportional to the temperature distance from the mean field critical temperature,
$\mathring{u}$ is positive. The properties of the random potential $V(x)$ are governed by a Gaussian
distribution with width $\mathring{\Delta}$ ($<<V(x)V(x')>>=4\mathring{\Delta}\delta(x-x')$,
the double angular brackets means averaging over disorder).  If $\mathring{\gamma}\equiv 0$
equations (\ref{eq_mov2}) describes dynamical properties
of a purely relaxational model (model A)  in the presence of disorder \cite{Lawrie84}.

We treat the critical dynamics of the disordered models within the field
theoretical RG method  \cite{Bausch76}, where the appropriate Lagrangians of the
models are studied. The average over the random potential generates new terms
in the Lagrangians with coupling $\mathring{\Delta}$, which correspond to the
static coupling terms in ${\mathcal H}$ generated by the disorder. The
renormalization of the Lagrangian leads to the RG functions, describing the
critical dynamics of our models. We use the minimal subtraction scheme with dimensional
regularization to calculate these functions.

For renormalization of the OP $\vec{\varphi}_0$,
fourth-order couplings $\mathring{u},\,\mathring{\Delta}$ and correlation functions with
$\vec{\varphi}^2_0$ insertion we introduce renormalization $Z$-factors as $\vec{\varphi}_0=Z^{1/2}_{\varphi}\vec{\varphi}$,
${\mathring u}=\mu^\epsilon Z_{\varphi}^{-2}Z_{u}u A_d^{-1}$,
${\mathring \Delta}=\mu^\epsilon Z_{\varphi}^{-2}Z_{\Delta}\Delta A_d^{-1}$,
and $|\vec{\varphi}_0|^2= Z_{\varphi^2}|\vec{\varphi}|^2$
($\mu$ is the scale, $\epsilon=4-d$ and $A_d$ is a geometric factor).
%Within the minimal subtraction approach $\mathring r$ renormalizes as
%${\mathring r}=Z_{\varphi^{2}} r$.
Within dynamics  renormalization factor for
the OP kinetic coefficient $\mathring \Gamma=Z_\Gamma\Gamma$ is introduced.
The $Z$-factors introduced so far are enough to renormalize the diluted
model A. For model C one needs to introduce additional
\begin{figure} \centering
\epsfig{file=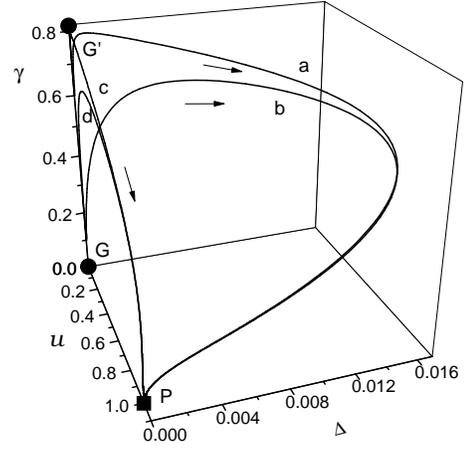,width=6.0cm,angle=0}
\caption{\label{n3statflow} Two loop flow of static parameters for $n=3$.
The stable fixed point (full square) has a zero width $\Delta$ of the impurity distribution
and leads to asymptotic exponents equal to the pure case (unstable fixed points are
indicated by full circles). However depending on the initial values of
the renormalized couplings the effective exponents are quite different and also depend on the inital
value of the width $\Delta$.}
\end{figure}
renormalization factors. The secondary density $m_0$ and coupling
parameter $\mathring{\gamma}$ are renormalized by $a_{m}^{1/2}m_0=Z_m m$ and
$a_{m}^{-1/2}{\mathring\gamma}=\mu^{\epsilon/2}Z_{\varphi^2}Z_m\gamma
A_d^{-1/2}$ with
$Z_{m}^{-2}(u,\Delta,\gamma)=1+\gamma^2 A_{\varphi^2}(u,\Delta)$.
The kinetic coefficient $\lambda$ renormalizes as
$a_{m}{\mathring \lambda}=Z^2_{m}\lambda$.

Defining the $\zeta$-functions as $d \ln Z^{-1}/ d\ln\mu$, where $Z$ represents
any renormalization factor, one obtains the flow equations for the renormalized static and
dynamic parameters. The flow equations for $u$ and $\Delta$ decouple from the
remaining parameters and are equal to expressions obtained for any $n$ in the diluted
Ginsburg-Landau-Wilson (GLW) model \cite{statics}.
For the additional static parameter $\gamma$ appearing in diluted model C we
have
\begin{equation}\label{dgamdl}
l\frac{d\gamma}{dl}=\gamma\Big[-\frac{\epsilon}{2}+\zeta_{\varphi^2}(u,\Delta)
+\frac{1}{2}\gamma^2B_{\varphi^2}(u,\Delta) \Big] \, .
\end{equation}
The flow parameter $l$ is related to the reduced temperature and is
consequently a measure for the distance to the critical temperature.
The function $\zeta_{\varphi^2}(u,\Delta)$ is known from statics in the diluted model \cite{statics}.
The function $B_{\varphi^2}(u,\Delta)$, which is defined by the additive
renormalization of the specific heat within the GLW-model, in the diluted
case reads $B_{\varphi^2}(u,\Delta)=n/2+{\cal O}(u^2,\Delta^2,u\Delta)$.

The flow equation for the time scale ratio (we introduce $\rho=w/(1+w)$ instead) is
\begin{equation}\label{dwdl}
l\frac{d\rho}{dl}=\rho (1-\rho)\Big[\zeta_{\Gamma}(u,\Delta,\gamma,\rho)-
\gamma^2B_{\varphi^2}(u,\Delta) \Big] \, ,
\end{equation}
where the dynamic $\zeta$-function $\zeta_{\Gamma}$ in two loop order reads
\begin{eqnarray} \label{Gamma}
\!\!&&\!\!\zeta_{\Gamma}(u,\Delta,\gamma,\rho)=\zeta_{\Gamma}^{(C)}(u,\gamma,\rho)+4{\Delta}-
\frac{n+2}{3}u\Delta+20\Delta^2 \nonumber\\
\!\!&&\!\!+2\Delta\rho\gamma^2\Bigg[3\Big(1-\ln(1-\rho)\Big)
+\rho\ln \frac{\rho}{1-\rho}-\frac{\rho}{1-\rho}\ln\rho\Bigg]   \, . \nonumber \\
\end{eqnarray}
The corresponding explicit two loop expression for the
$\zeta$-function of model C, $\zeta_{\Gamma}^{(C)}(u,\gamma,\rho)=
\zeta_w^{(C)}(u,\gamma,\rho)+n\gamma^2/2$, is given in \cite{Folk04}
(see $\zeta_w$ in Eq. (50) there). In these terms also the pure model A terms are included, which
taken together with the last three terms in the first line of Eq. (\ref{Gamma}) recover the
 $\zeta$-function for the diluted model A \cite{Lawrie84,Prudnikov92}.
The zeros of the right hand sides of Eqs.(\ref{dgamdl}), (\ref{dwdl}), and the
corresponding equations for $u$ and $\Delta$ give the possible fixed points.
\begin{table}[t]
\begin{ruledtabular}
\caption {\label{tab1} Static and dynamic fixed points for $n=1$. Fixed points
with $\rho^*=1$ ($w^*=\infty$) are always unstable and not shown. {\bf M} is the stable
fixed point of the diluted model C.}
\begin{tabular}{ccccc}
 FP & $u^*$ & $\Delta^*$ & $\gamma^*$&$\rho^*$\\  \hline
{\bf G} & 0&0 &0 &0\\
{\bf G}$^{\prime}$ &0 &0 & 1.414 &0\\
{\bf P}&1.315 &0 &0&0\\
{\bf P}$^{\prime}$&1.315 &0 &0.458&0 \\
{\bf C}& 1.315&0 &0.458&0.266 \\
{\bf M} & 1.633 &0.021 &0&0
\end{tabular}
\end{ruledtabular}
\end{table}
\begin{table}[b]
\begin{ruledtabular}
\caption {\label{tab2} Stability exponents according to the fixed points in Tab.\ref{tab1}
for $n=1$.}
   \begin{tabular}{ccccc}
FP&$\omega_u$ & $\omega_{\Delta}$ &$\omega_{\gamma}$&$\omega_{\rho}$\\   \hline
{\bf G}& -1&-1&-0.5&0\\
{\bf G}$^{\prime}$& -1&-1&1&0\\
{\bf P}&0.566 &-0.105&-0.053 &0.052\\
{\bf P}$^{\prime}$&0.566 &-0.105 &0.105&-0.053 \\
{\bf C}&0.566&-0.105&0.105&0.041 \\
{\bf M}&0.494 &0.194 &0.0018&1.139
\end{tabular}
\end{ruledtabular}
\end{table}

It turns out that for all values of $n$ the stable fixed point value $\gamma^\star=0$, and the well-known
asymptotic static results are reproduced. Thus the flow in the space of the static
couplings $u$, $\Delta$ and $\gamma$ for $n<n_c<2$ looks like the flow for $n=1$ (see Fig. \ref{n1statflow})
whereas for $n>n_c$ it looks like the flow for $n=3$ (see Fig. \ref{n3statflow}). The fixed points for $n=1$
are indicated in Tab. \ref{tab1}. Only the mixed fixed point ({\bf M} in Fig.\ref{n1statflow}) is stable.
However depending on the initial conditions a rich crossover behavior is observed.
The same is true for  $n=3$ although now the pure fixed point ({\bf P} in Fig. \ref{n3statflow}) is stable.

The stable fixed points are found by calculating the stability matrix  and its eigenvalues $\omega_i$
($i$ represents $u$, $\Delta$, $\gamma$ or $\rho$) . They govern the ''velocity'' of the flow near the fixed points. A small stability exponent
at the stable fixed point indicates a slow approach of the asymptotic  behavior. This is the case for all
$n$ and can be seen from Tab.\ref{tab2} for $n=1$. It is the slow approach in the $\gamma$ direction which
characterizes the static flow within model C (see the extremely small values $\omega_\gamma=0.0018$ for $n=1$
[at the fixed point {\bf M} in Tab.\ref{tab2}]  and for $n=3$ still $\omega_\gamma=0.1109$).
In addition to the small values of the static stability exponents near the fixed points one has also
small values $\omega_\rho$ coming from the dynamic parameter $\rho$  (however only near unstable fixed points).
Thus one expects depending on the initial values of static and dynamic parameters a complex behavior in the
non-asymptotic region.
\begin{figure} \centering
\epsfig{file=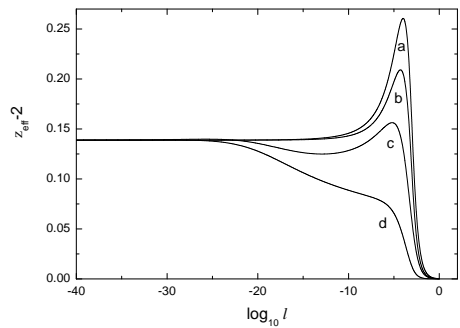,width=6.0cm,angle=0}
\caption{\label{n1zeff} Effective dynamical exponent for $n=1$ for different initial conditions. The lines
correspond to the static flows shown in Fig.\ref{n1statflow}.}
\end{figure}
\begin{figure}[b] \centering
\epsfig{file=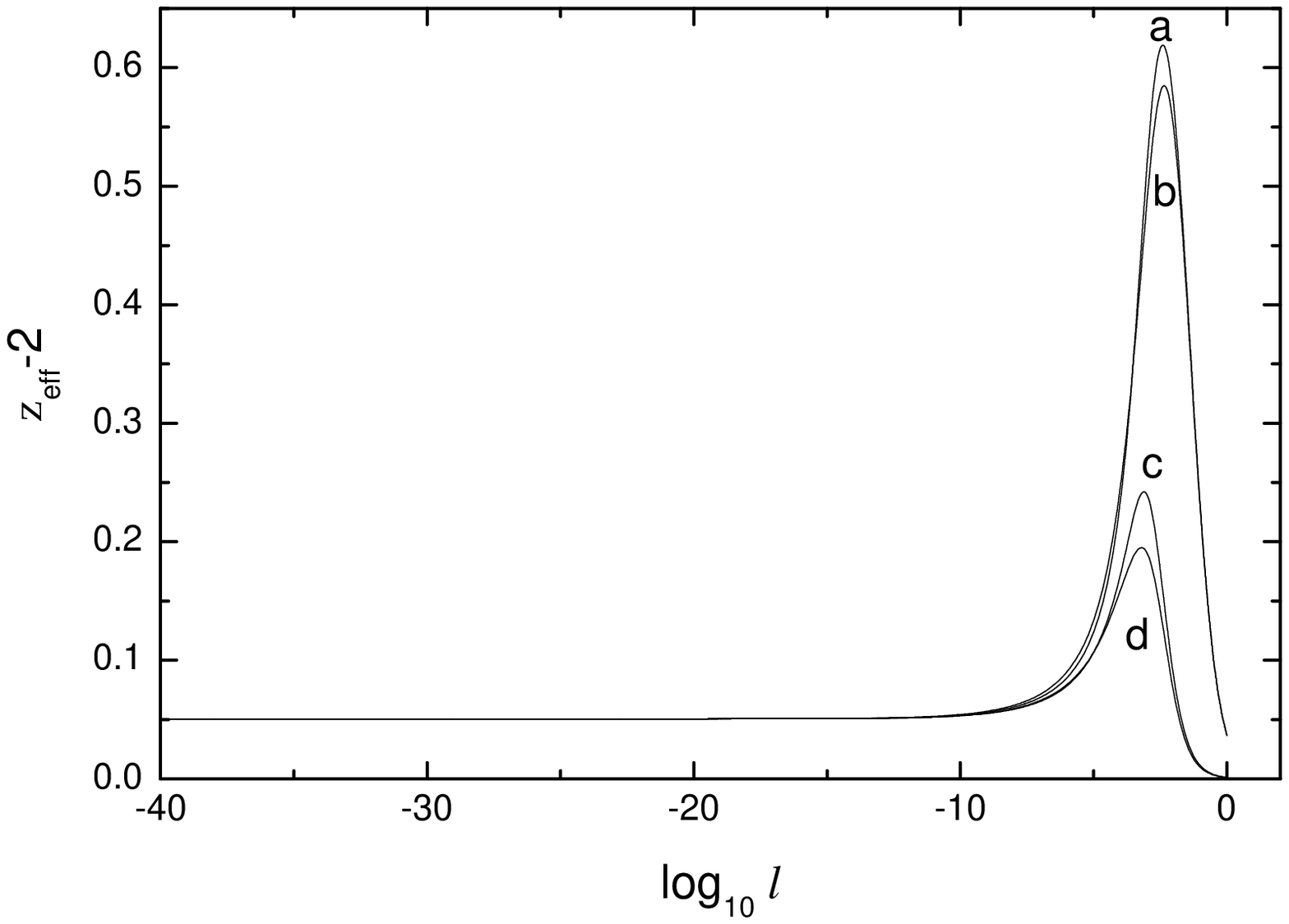,width=6.0cm,angle=0}
\caption{\label{n3zeff} Effective dynamical exponent for $n=3$ for different initial conditions. The lines
correspond to the static flows shown in Fig.\ref{n1statflow}.}
\end{figure}
The effective dynamical exponent is of special interest here, it is  found by inserting the solutions of the
flow equations into the expression
\begin{equation}
z_{eff}(l)=2+\zeta_{\Gamma}\big(u(l),\Delta(l),\gamma(l),\rho(l)\big)  \, .
\end{equation}
One reaches  the universal asymptotic value $z$ when the flow comes very near the stable fixed point.
One observes that the stable fixed point value of the time ratio is always zero,
independently of the specific heat exponent value of the pure model. Consequently, the results for
model A are recovered in the asymptotics (either in the diluted universality class for $n<n_c$ or in the pure model A
universality class for $n>n_c$). This combines with the fact that even in the region where
dilution changes the static critical behavior the stable fixed point value of $\gamma^*$ is always zero.
In consequence the secondary density is for all $n$ asymptotically decoupled and has an asymptotic
dynamical exponent $z_m=2$.

The non-asymptotic behavior is however quite different as can be seen from Figs. \ref{n1zeff} and \ref{n3zeff}.
Due to the static non-asymptotic behavior also the dynamics  is dominated by different non-asymptotic effects.
This can be seen by comparing different $z_{eff}(l)$ for different initial conditions. The curves a, b, and c in Fig.
\ref{n1statflow} and all curves in Fig. \ref{n3statflow} reach large values of the coupling $\gamma$ and/or $\Delta$
and this leads  to the typical maximum in the effective exponents independent of the initial value of $\rho$
(for the statics see e.g. \cite{perumal03}). However an additional fixed point {\bf P}$^\prime$ is present at $n=1$.
This leads for curve d in Fig. \ref{n1statflow} almost to a plateau of $z_{eff}$ at its value for the unstable fixed point
{\bf P}$^\prime$. This plateau is more pronounced when the flow comes nearer to  {\bf P}$^\prime$ where it stays longer because of the
small transient exponent $\omega_\rho$. For curve c both effects (the maximum and the effect of fixed point {\bf P}$^\prime$) are combined
leading to the minimum in $z_{eff}$.
\begin{figure} \centering
\epsfig{file=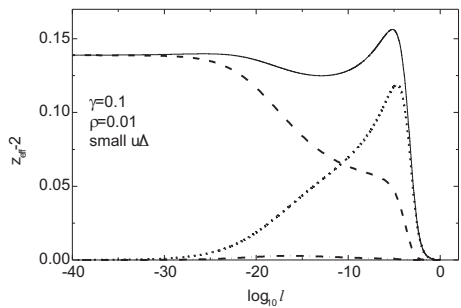,width=6.0cm,angle=0}
\caption{\label{n1contributions} Different contributions to the effective dynamical critical
exponent for $n=1$. The solid line represents the total $z_{eff}$, for the other lines see the text. }
\end{figure}
Consider now the contributions to the effective dynamical critical exponent $z_{eff}$ of different origin:
(i) from the terms already present in model A (dashed curve in Fig. \ref{n1contributions}), (ii) from terms present in
pure model C only (short dashed curve) and (iii) and from terms present in the diluted model C  only
(dashed-short dashed curve). The above contributions may add up to almost the {\bf asymptotic}
value of the exponent {\bf although} the parameters are far away form their asymptotic values. This is an important point since
the appearance of an asymptotic value in one physical quantity does not mean that other quantities have also reached the
asymptotics. This is due to the different dependence of physical quantities on the model parameters.
Another special feature of the diluted model C is that already in one loop order one observes qualitatively the same
behavior as in two loop order of course with changed values for the exponents and the borderline value $n_c$, which in one
loop is at $n_c=4$.

In concluding we remark that contrary to the general belief the coupling of a conserved density to the order
parameter is relevant for the calculation of the dynamical critical behavior  of diluted systems since this coupling is
important to describe non-asymptotic effects.  These effects have been  seen
in the experiments on physical systems \cite{perumal03,EX} as well as in Monte Carlo simulations \cite{MC}.

We acknowledge support from the Fonds zur F\"orderung der wissenschaftlichen Forschung
(project P16574)

 \end{document}